\documentclass[11pt]{article}

\usepackage[final]{acl}

\usepackage{times}
\usepackage{latexsym}

\usepackage[T1]{fontenc}
\usepackage[utf8]{inputenc}

\usepackage{microtype}

\usepackage{inconsolata}

\usepackage{graphicx}
\usepackage{booktabs}
\usepackage{hyperref}
\usepackage{amsmath}

\title{AstroMLab 5: Structured Summaries and Concept Extraction\\ for 400,000 Astrophysics Papers}

\author{
  \textbf{Yuan-Sen Ting}\textsuperscript{1},
  \textbf{Alberto Accomazzi}\textsuperscript{2},
  \textbf{Tirthankar Ghosal}\textsuperscript{3},\\
  \textbf{Tuan Dung Nguyen}\textsuperscript{4},
  \textbf{Rui Pan}\textsuperscript{5},
  \textbf{Zechang Sun}\textsuperscript{6},
  \textbf{Tijmen de Haan}\textsuperscript{7}
\\
\\
  \textsuperscript{1}The Ohio State University,
  \textsuperscript{2}Harvard-Smithsonian Center for Astrophysics,\\
  \textsuperscript{3}Oak Ridge National Laboratory,
  \textsuperscript{4}University of Pennsylvania,\\
  \textsuperscript{5}University of Illinois at Urbana-Champaign,
  \textsuperscript{6}Tsinghua University,
  \textsuperscript{7}KEK\\
}

\begin{document}
\maketitle

\begin{abstract}
We present a dataset of 408,590 astrophysics papers from arXiv (astro-ph), spanning 1992 through July 2025. Each paper has been processed through a multi-stage pipeline to produce: (1) structured summaries organized into six semantic sections (Background, Motivation, Methodology, Results, Interpretation, Implication), and (2) concept extraction yielding 9,999 unique concepts with detailed descriptions. The dataset contains 3.8 million paper-concept associations and includes semantic embeddings for all concepts. Comparison with traditional ADS keywords reveals that the concepts provide denser coverage and more uniform distribution, while analysis of embedding space structure demonstrates that concepts are semantically dispersed within papers—enabling discovery through multiple diverse entry points. Concept vocabulary and embeddings are publicly released at \url{https://github.com/tingyuansen/astro-ph_knowledge_graph}.
\end{abstract}

\section{Introduction}

A frontier application of large language models is their deployment as autonomous agents that reason about scientific literature, plan research strategies, and execute multi-step retrieval tasks~\citep{brown2020language, Wang2025b}. Such systems, already demonstrated in materials science and chemistry for autonomous experimentation~\citep{Szymanski2023, Boiko2023, Bran2023, Ramos2024}, require structured knowledge representations to function—moving beyond language processing to operate on semantically organized information. While LLMs can process raw text, their utility as research agents depends on access to curated intermediate representations that bridge unstructured documents and formal knowledge structures~\citep{Lewis2020}.

Astronomy presents an advantageous testing ground: most papers are archived on arXiv (astro-ph since 1992), and the open-sky policy enables databases to link astronomical objects directly to papers. The combination of papers, observed objects, and their properties provides an ecosystem where structured representations could enable agentic research and autonomous discovery.

However, text as a modality remains under-curated in the astronomy literature. Current resources are either too complete (full source, which is difficult to extract insights) or too sparse (abstracts only). Both extremes limit downstream applications. Useful scientific ideas emerge from holistic understanding of concepts rather than direct processing of individual words—this is how humans engage with literature. Keywords were designed to bridge this gap, but when present, are rarely mapped to controlled vocabularies like the Unified Astronomy Thesaurus (UAT) and exhibit sparsity—most keywords appear in very few papers and many papers have very few keywords, rendering them unsuitable for systematic analysis.

LLMs can extract such structured representations from papers, but this is cost-intensive. Individual researchers performing this task separately would waste computational resources. A single, centralized effort provides economies of scale and ensures consistency across the literature.

To address this, we organize all astro-ph papers into structured summaries and concepts—two intermediate layers that bridge the gap between raw text and knowledge representation. Our work builds on recent developments in applying LLMs to astronomical research, including domain-specific models like AstroLLaMA~\citep{Pan2024} and AstroSage~\citep{deHaan2025a}, complementary efforts in knowledge graph construction~\citep{Sun2024, Kau2024}, and the development of recommender systems~\citep{Geng2022, Chu2023, Zhao2023, Vats2024}. We present a comprehensive dataset spanning 408,590 papers with 9,999 unique concepts, their semantic embeddings, and structured summaries. 

\section{PDF to Text: OCR Pipeline}

Converting astrophysics PDFs to machine-readable text presents challenges due to the prevalence of mathematical equations, multi-column layouts, and figures with embedded captions. We chose to use OCR rather than LaTeX source files because LaTeX sources are not uniformly structured across papers and many papers use custom macros that complicate parsing.

Our pipeline initially used Nougat~\citep{blecher2023nougat}\footnote{\url{https://github.com/facebookresearch/nougat}}, an academic document OCR model that converts PDFs to markdown format for ease of processing. Processing each paper requires approximately 1 minute on a V100 GPU, representing a substantial computational investment—processing over 350,000 papers required about 6,000 V100 GPU-hours. Starting in November 2024, we transitioned to Mathpix OCR API\footnote{\url{https://mathpix.com/ocr}} as it proved more reliable than Nougat.

Both Nougat and Mathpix preserve mathematical notation in LaTeX format within the OCR output. For section detection, Nougat outputs markdown headers (\texttt{\#\#\#}) while Mathpix preserves LaTeX section commands (\texttt{\textbackslash section}). The transition to Mathpix was motivated by Nougat's occasional failure mode where approximately 1 in 500 pages would produce repetitive text; such corrupted pages are naturally excluded during the summarization stage, though this may result in missing information at a subdominant level. For Mathpix OCR (covering approximately 50,000 papers from November 2024 onward), author team inspection of randomly sampled pages revealed no systematic OCR errors at levels that would impact summary quality.

\section{Multi-Stage Summarization}

\subsection{Chunk-Based Compression}

During early development, we found that processing entire papers at once led to incomplete summaries, with LLMs often omitting important details or providing superficial coverage. This was problematic for generating organized summaries with properly populated sections—methodological details, for instance, were frequently under-represented. Processing single abstracts typically missed useful information like detailed derivations and technical implementation specifics. This motivated our chunk-based approach, which processes papers in manageable segments while maintaining context across chunks. 

We split each paper into approximately 10,000-character chunks using section-aware boundaries to avoid mid-sentence breaks. Papers are split at section boundaries (detecting either markdown headers from Nougat or \texttt{\textbackslash section} commands from Mathpix), with adjacent small sections merged up to the 10,000-character limit. Each chunk is sequentially compressed with context from previously compressed chunks, ensuring coherence across the full paper. This approach increases token costs several-fold compared to single-pass processing, but when this project started in late 2023, this was necessary to achieve adequate quality. We maintained this approach for subsequently processed papers to ensure consistency.

The compression system prompt emphasizes: (1) retaining LaTeX formulas, (2) focusing on motivations and methods, (3) highlighting key results and connections to other works, (4) preserving technical jargon for expert readers, and (5) excluding acknowledgments and references. As language models improved, we adopted the most affordable versions while maintaining quality. Different papers were processed with GPT-4o, GPT-4o-mini, o1-mini, and DeepSeek-v3 depending on availability. The complete summarization process for all 408,590 papers required over \$50,000 in API costs, not including OCR costs.

\subsection{From Raw Summaries to Structured Organization}

Abstract sections often jumble information chronologically or by importance, making systematic analysis difficult. We reorganize raw summaries into seven semantic sections that follow the logical flow of scientific papers: Title and Author, Background, Motivation, Methodology, Results, Interpretation, and Implication. This structured format enables targeted queries and facilitates knowledge representation by clearly separating context, methods, and outcomes. Appendix~\ref{app:example_summary} shows a complete example demonstrating all six sections.

\section{Concept Extraction and Vocabulary}

\subsection{Extraction Methodology}

For each organized summary, we prompt the LLM to extract approximately 10 key concepts focusing on novel contributions. The target of 10 concepts provides finer granularity than traditional keyword systems (where author-supplied keywords typically number 3-5 per paper) while remaining tractable for LLM extraction. The system prompt emphasizes: (1) identifying innovations and novel methods, (2) covering both scientific concepts (observational phenomena, theoretical frameworks) and technological concepts (computational techniques, instrumentation), and (3) avoiding generic field names or overly specific parameters. 

This approach leverages the capacity of modern language models~\citep{achiam2023gpt, Beltagy2019,Ting2025} to understand domain-specific scientific contexts. Each concept includes three components: a \textbf{Name} (3-4 word concise label), a \textbf{Class} (Cosmology \& Nongalactic Physics, High Energy Astrophysics, Instrumental Design, Galaxy Physics, Numerical Simulation, Statistics \& AI, Solar \& Stellar Physics, or Earth \& Planetary Science), and a \textbf{Description} ($\sim$100-word technical explanation). The final concept vocabulary was generated homogeneously using a combination of GPT-4o and o1-mini to ensure consistency across all papers.

\subsection{Vocabulary Construction and Clustering}

LLM-extracted concepts lack a priori control over consistency—different papers may use different terminology for the same concept, and there is no guarantee of controlled vocabulary. To address this, following the methodology of \citet{Sun2024}, we employ a multi-stage clustering process. For each extracted concept in each paper, we combine the organized summary with the concept name to generate detailed descriptions. These descriptions are then embedded using OpenAI's text-embedding-3-large model. We perform K-means clustering (k=10,000) in the cosine similarity space to consolidate similar concepts, merging semantically equivalent variants into single unified entries. The clustering maximizes inter-cluster distances while grouping semantically similar extractions. The clustering process synthesizes new unified concept descriptions that capture the full semantic range across papers.

We experimented with different vocabulary granularities in log space (3,000, 10,000, and 30,000 concepts). We found 10,000 concepts to provide the most useful balance. All 10,000 concepts and descriptions were manually reviewed by the author team, during which one null concept (representing rare OCR failure cases) was identified and removed, leaving the final vocabulary of 9,999. The concepts have been used in various downstream analyses (Sections 5-6) providing ongoing validation. Given the dataset scale, our validation strategy prioritized full vocabulary review over exhaustive paper-by-paper evaluation, and users should exercise appropriate scrutiny when employing the dataset for specific applications.

Each concept retains its detailed description synthesized from multiple papers, providing more context than typical keyword systems. The concept distribution across categories is shown in Table~\ref{tab:categories}. Each concept appears in an average of 383 papers (median: 223), making them statistically robust while maintaining sufficient specificity. As we will see in Section~\ref{sec:keywords}, this granularity avoids both the overly broad categories and overly specific identifiers that plague traditional keyword systems.
\begin{table}[t]
  \centering
\small
\begin{tabular}{lr}
\toprule
\textbf{Category} & \textbf{Count} \\
\midrule
Cosmology \& Nongalactic Physics & 2,192 \\
High Energy Astrophysics & 1,606 \\
Instrumental Design & 1,295 \\
Galaxy Physics & 1,267 \\
Numerical Simulation & 1,050 \\
Statistics \& AI & 1,020 \\
Solar \& Stellar Physics & 930 \\
Earth \& Planetary Science & 639 \\
\bottomrule
  \end{tabular}
\caption{Distribution of 9,999 concepts across research categories.}
\label{tab:categories}
\end{table}

\section{Quality Evaluation}
\label{sec:keywords}

\subsection{Comparison with Traditional Keywords}

To evaluate our concept vocabulary, we compare it with traditional ADS keywords extracted via the NASA ADS API for all 408,590 papers. ADS keywords are author-supplied and not systematically checked against controlled vocabularies like the Unified Astronomy Thesaurus. We performed curation by removing arXiv classification keywords (e.g., "Astrophysics - Cosmology"), normalizing to lowercase, and filtering overly common keywords (>20,000 occurrences) and rare keywords (<10 occurrences). After curation, ADS keywords cover 73\% of papers (298,658) with 6,909 unique keywords and 1.27M associations.

\begin{figure*}[t]
\centering
\includegraphics[width=0.95\textwidth]{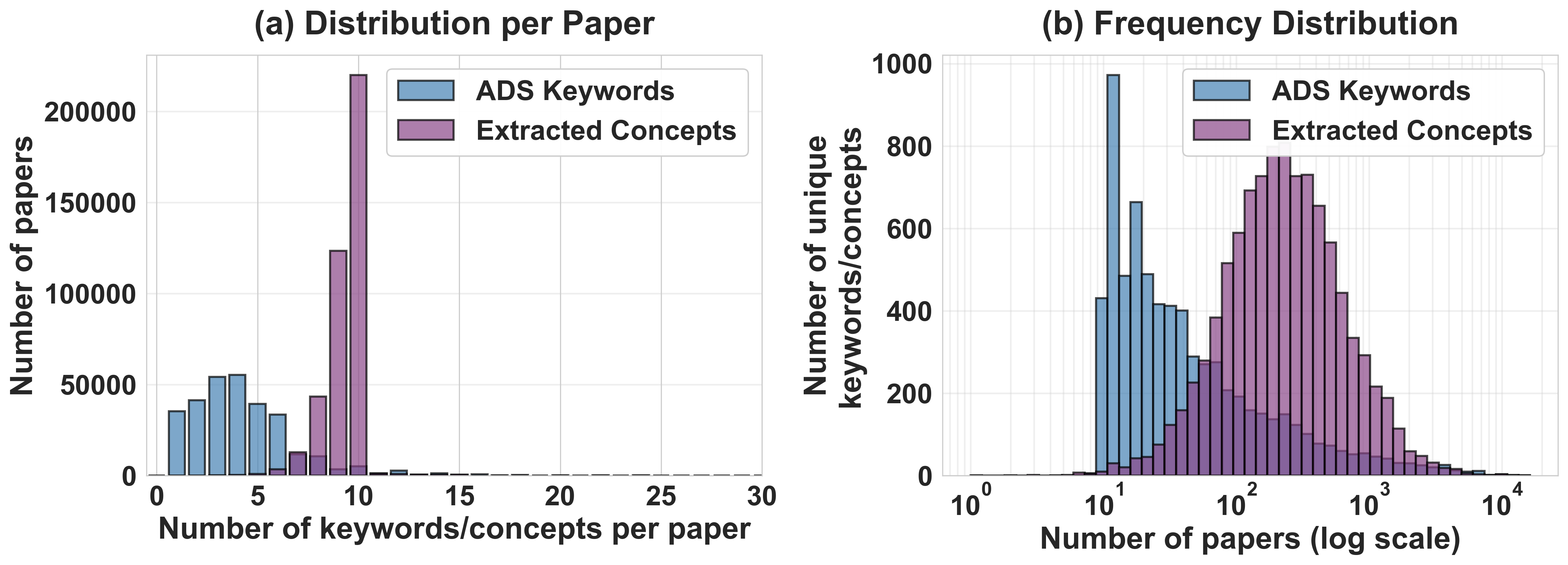}
\caption{Distribution of keywords/concepts per paper (left) and frequency distribution (right). ADS keywords show high sparsity with many papers having few keywords, while our concepts provide consistent coverage. The frequency distribution (right) reveals a large pile-up of overly generic terms and an extended tail of overly specific identifiers, while our concepts maintain more balanced intermediate granularity.}
\label{fig:keyword_comparison}
\end{figure*}

Figure~\ref{fig:keyword_comparison} shows two key differences. First, ADS keywords suffer from severe sparsity: 44\% of papers have $\leq$3 keywords and 62\% have $\leq$4 keywords—insufficient for effective semantic search or recommendation systems. This primarily reflects different generation mechanisms: author-supplied keywords are known to be sparse partly because authors often do not systematically check controlled vocabularies, and different journals have varying keyword standards. In contrast, our extraction prompt explicitly requests approximately 10 concepts per paper, achieving complete coverage: all 408,590 papers have structured summaries and concept associations. The average paper has 9.4 concepts (median: 10) concepts per paper with a small dispersion.

Beyond coverage, our concepts exhibit more balanced frequency distribution. While ADS keywords suffer from extreme imbalances—most common keywords like "galaxies: evolution" appear in 16,321 papers while 2,658 keywords (38\% of the vocabulary) appear in only 10-20 papers—our concepts maintain intermediate granularity. Table~\ref{tab:keyword_problems} illustrates this problem: the most common keywords represent overly broad field categories with limited discriminative power, while rare keywords are often object-specific identifiers (e.g., "grb 080319b") rather than research themes.

\begin{table}[t]
\centering
\small
\begin{tabular}{lrr}
\toprule
\textbf{Keyword} & \textbf{Papers} & \textbf{\%} \\
\midrule
\multicolumn{3}{c}{\textit{Most Common (Overly Broad)}} \\
\midrule
galaxies: evolution & 16,321 & 5.5 \\
galaxies: active & 14,121 & 4.7 \\
accretion & 12,540 & 4.2 \\
methods: numerical & 12,510 & 4.2 \\
stars: formation & 9,172 & 3.1 \\
dark matter & 9,032 & 3.0 \\
methods: data analysis & 8,384 & 2.8 \\
galaxies: formation & 8,313 & 2.8 \\
\midrule
\multicolumn{3}{c}{\textit{Rare (Overly Specific) - 2,658 keywords with 10-20 papers}} \\
\midrule
\multicolumn{3}{l}{gamma-ray burst: individual: grb 080319b} \\
\multicolumn{3}{l}{stars: individual: alphanumeric: hd 209458} \\
\multicolumn{3}{l}{galaxies: individual: alphanumeric: ngc 1275} \\
\multicolumn{3}{l}{pulsars: individual: alphanumeric: psr j1614-2230} \\
\multicolumn{3}{l}{x-rays: binaries: individual: alphanumeric: cygnus x-1} \\
\bottomrule
  \end{tabular}
\caption{Examples of overly broad and overly specific ADS keywords. The most common keywords represent broad field categories with limited discriminative power, while rare keywords are often object-specific with minimal value for thematic analysis.}
\label{tab:keyword_problems}
\end{table}

In contrast, our concepts balance these extremes through systematic curation and maintain semantic meaningfulness across all frequency ranges, through the clustering and consolidation process. Table~\ref{tab:concept_examples} demonstrates this: high-frequency concepts represent general methodologies applicable across subfields (e.g., "Monte Carlo Simulations" with 13,671 papers) that retain semantic specificity and discriminative power for retrieval, medium-frequency concepts capture well-established research areas (e.g., "Stellar Evolution Models" with 2,751 papers), and low-frequency concepts identify emerging or specialized topics while remaining thematic rather than object-specific (e.g., "Interpretable Machine Learning in Astronomy" with 49 papers). The median concept appears in 223 papers—more balanced than the median of 28 for ADS keywords.

\begin{table*}[t]
\centering
\small
\begin{tabular}{llrl}
\toprule
\textbf{Frequency} & \textbf{Concept} & \textbf{Papers} & \textbf{Class} \\
\midrule
High & Monte Carlo Simulations & 13,671 & Numerical Simulation \\
 & N-Body Simulation Dynamics & 12,041 & Numerical Simulation \\
 & Astronomical Spectral Energy Profiles & 9,590 & Galaxy Physics \\
 & Cosmic Microwave Background & 9,166 & Cosmology \& Nongalactic Physics \\
\midrule
Medium-High & Stellar Evolution Models & 2,751 & Solar \& Stellar Physics \\
 & Dynamic Cosmological Constant & 2,692 & Cosmology \& Nongalactic Physics \\
 & Galaxy Morphological Study & 2,485 & Galaxy Physics \\
 & High-Redshift Quasars & 1,626 & Cosmology \& Nongalactic Physics \\
\midrule
Medium & Gravitational Lensing Surveys & 278 & Instrumental Design \\
 & CMB Simulation Methodologies & 276 & Numerical Simulation \\
 & Marginalization in Bayesian Inference & 268 & Statistics \& AI \\
 & Neural Inference Methods & 202 & Statistics \& AI \\
\midrule
Low & Interpretable Machine Learning in Astronomy & 49 & Statistics \& AI \\
 & Neutrino-Driven Supernova Simulations & 48 & Numerical Simulation \\
 & Exoplanetary Companion Systems & 44 & Solar \& Stellar Physics \\
 & Gravitational Wavefront Interactions & 35 & Cosmology \& Nongalactic Physics \\
\bottomrule
\end{tabular}
\caption{Examples of our concepts across frequency ranges. Unlike traditional keywords that become meaningless at extremes (overly broad or object-specific), our concepts remain scientifically meaningful across all frequencies, maintaining thematic coherence rather than becoming object-specific identifiers.}
\label{tab:concept_examples}
\end{table*}

These limitations of traditional keywords make content-based recommendation systems difficult to implement. This dataset provides an alternative that enables more robust semantic search and recommendation algorithms, which may be useful for platforms like NASA ADS. Our concept vocabulary includes many terms not present in the UAT, including in emerging areas like deep learning applications in astronomy. While our concepts can be mapped to UAT for compatibility with existing systems, we also propose this vocabulary as a potential foundation for extending or complementing the UAT with contemporary research terminology.

\begin{figure*}[ht]
  \centering
  \includegraphics[width=0.9\textwidth]{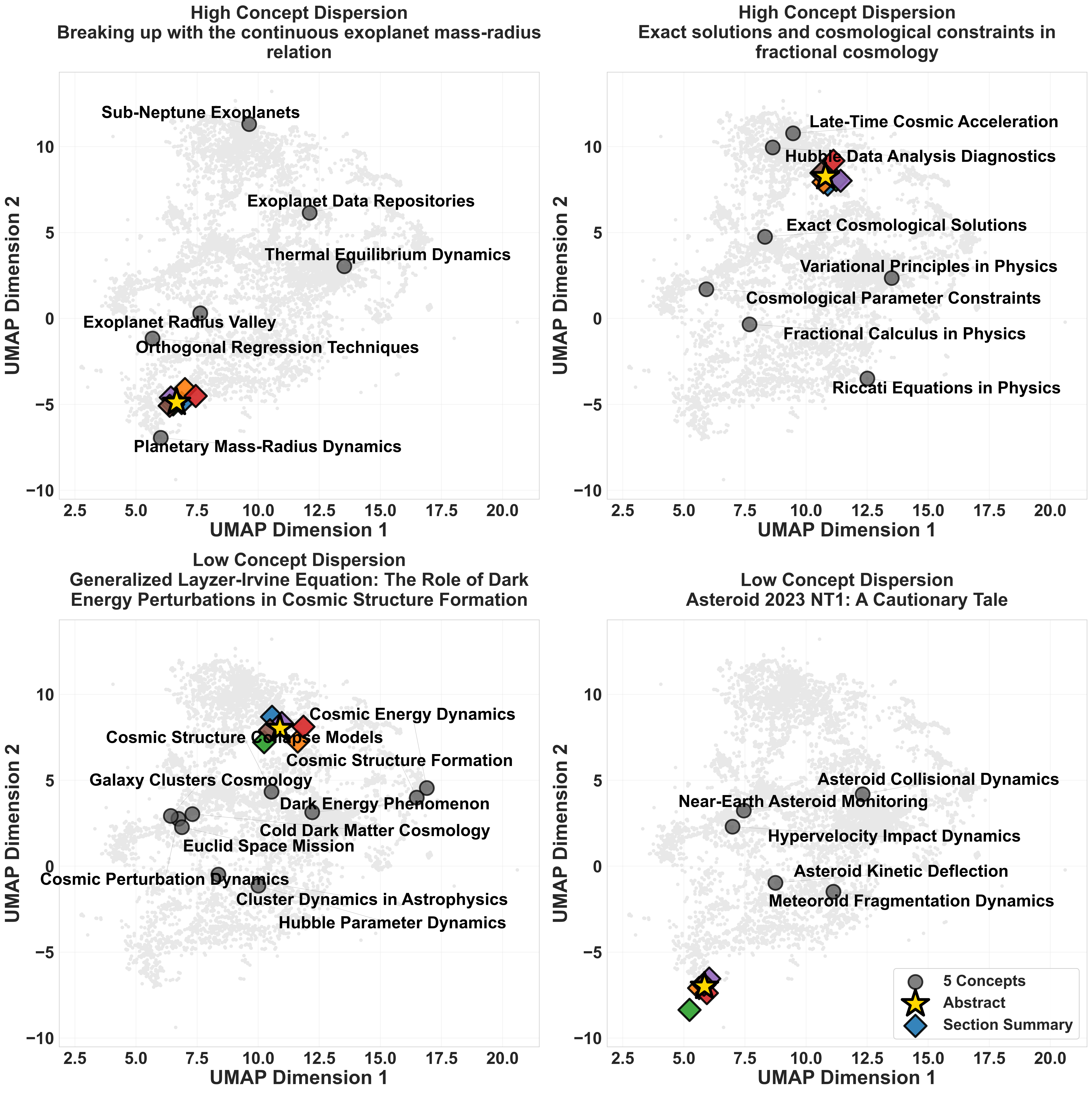}
  \caption{UMAP projections of concept (grey symbols) and summary embeddings (colored diamonds) and the abstract (gold star) for four representative papers. Faint gray background shows all 9,999 concepts in the vocabulary. Even papers classified as "low dispersion" (bottom row) have concepts spread across distinct semantic regions, showing that abstracts (and summaries) cannot capture the full conceptual diversity present in papers, unlike concepts.}
  \label{fig:umap_examples}
\end{figure*}

\subsection{Concepts for Discovery}

Beyond coverage and frequency balance, why are concepts superior to abstracts for discovery tasks? While abstracts provide summaries of papers, they operate at a narrative level that is not optimal for discovery. Novel ideas often emerge from specific methodological details, intermediate results, or conceptual connections that are embedded within a paper but not prominently featured in its abstract. Furthermore, current language models process continuous text rather than discrete conceptual tokens, limiting their ability to generate novel hypotheses through systematic exploration of the idea space.

To demonstrate why concepts are critical for discovery, we analyze the embedding space structure of 10,000 randomly sampled papers. Each concept in our vocabulary has a detailed description (see Table~\ref{tab:concept_examples}), from which we extract embeddings using OpenAI's text-embedding-3-large. We perform the same embedding extraction for each paper's abstract and for the six individual sections of its structured summary.

Figure~\ref{fig:umap_examples} shows UMAP projections of four representative papers—two with high concept dispersion (top row) and two with low dispersion (bottom row). The faint gray background represents all 9,999 concepts in our vocabulary, providing spatial context. Even in cases labeled as "low dispersion" (bottom row), the concepts assigned to individual papers (bold gray circles with labels) remain dispersed across semantic space. 

This dispersion occurs because papers contain multiple distinct ideas spanning different domains. For example, the top-right panel shows a paper on fractional cosmology that discusses concepts including "Hubble Data Analysis Diagnostics", "Variational Principles in Physics", "Fractional Calculus in Physics", and "Riccati Equations in Physics"—concepts that occupy distant regions of semantic space, bridging observational analysis, theoretical cosmology, and mathematical physics. Such concepts cannot be recovered from abstract embeddings alone; some are embedded deeply in methodological sections and never explicitly mentioned in abstracts. In stark contrast, the six summary sections (colored diamonds) and abstract (gold star) cluster tightly together in all cases, as all sections describe the same paper from different angles—they are semantically coherent because they narrate a single research story.

This analysis does not diminish the value of structured summaries—quite the contrary. It reveals the complementarity of concepts and summaries in our knowledge graph. Concepts are dispersed across semantic space, assigned to papers based on diverse topical content, making them ideal for discovery. A researcher exploring "Variational Principles in Physics" can find relevant papers, even if this concept appears only in a methodological subsection and not in the abstract. Summaries, conversely, cluster together because all sections describe the same paper. This narrative coherence makes them valuable for understanding context after relevant papers are identified through concept-based discovery.

\begin{figure*}[!htbp]
  \centering
  \includegraphics[width=0.95\textwidth]{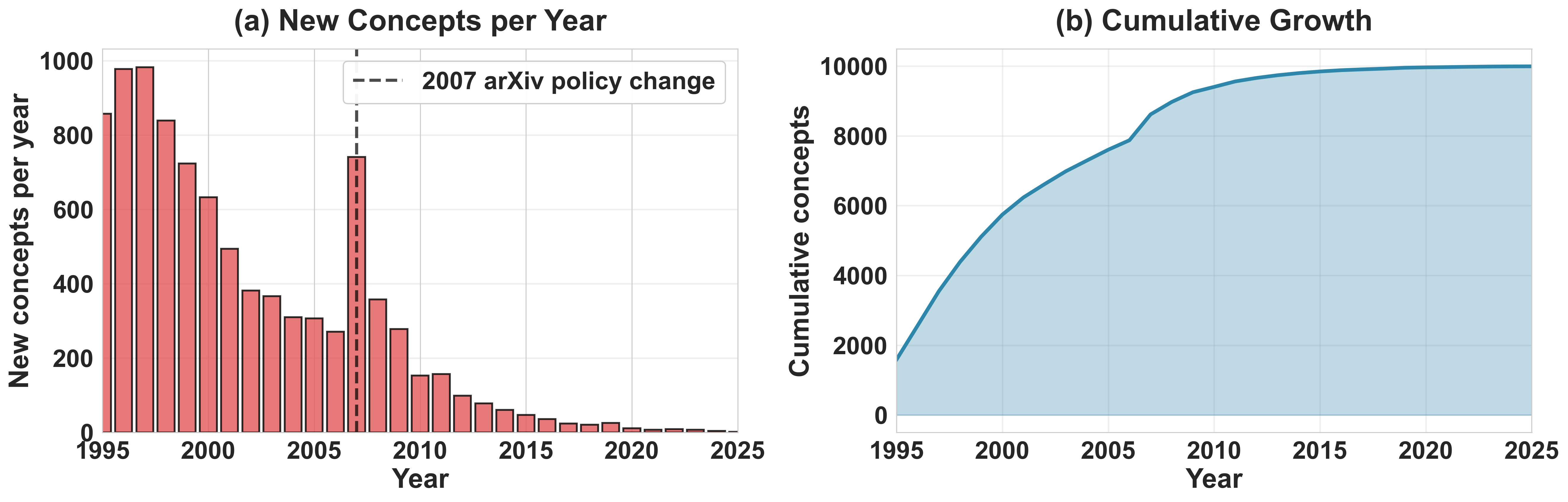}
  \caption{Temporal evolution of concept vocabulary across three decades. (a) Number of new concepts emerging each year (crossing the 5-paper threshold). (b) Cumulative growth of the concept vocabulary. The rapid expansion in the early years reflects foundational concepts when arXiv began. A secondary peak in 2007 corresponds to cross-listing policy changes.}
  \label{fig:concept_emergence}
  \end{figure*}
  
\section{Applications}

Having established the quality advantages of our concept vocabulary, we now demonstrate its utility through two applications that leverage these properties: temporal analysis of concept emergence and co-occurrence analysis of research themes.

\subsection{Temporal Evolution of Concepts}

The granular and semantically meaningful nature of our concepts enables precise tracking of how ideas emerge, evolve, and connect across different research areas. This application demonstrates the value of our vocabulary for constructing knowledge graphs~\citep{Kau2024} that trace research evolution. We analyze concept emergence by identifying when each concept first appeared in at least 5 papers (a threshold ensuring stability rather than single-paper anomalies). Figure~\ref{fig:concept_emergence} shows the temporal evolution across three decades, with new concepts per year (left) and cumulative growth (right).

The declining rate of new concept emergence in recent years does not necessarily indicate reduced innovation. Several factors contribute to this pattern. First, many fields have matured, with research increasingly focused on connections between established concepts rather than entirely new topics. Second, our 5-paper threshold means concepts can appear earlier than their peak importance—for example, concepts about the Gaia mission and the James Webbs Space Telescope emerged earlier than their launch, when early planning papers crossed the threshold, despite these missions becoming prominent only after launch. Third, our clustering methodology itself may exhibit systematic bias: because clustering aims to consolidate semantically similar terms across all papers, genuinely novel concepts appearing in recent years may be merged into established clusters from earlier periods if sufficiently similar in embedding space, suppressing the apparent emergence rate.

A notable secondary peak occurred in 2007, corresponding to arXiv expanding cross-listing policies to allow papers from other disciplines to include astro-ph as a secondary category. Analysis of these 2007 concepts reveals their origin: 51\% are classified as Cosmology \& Nongalactic Physics and 13\% as High Energy Astrophysics, dominated by theoretical topics including Loop Quantum Gravity, Holographic Duality, Einstein-Gauss-Bonnet Gravity Theories, Quantum Entanglement Entropy, Conformal Field Theory, and Type IIB and Heterotic String Theories. These reflect contributions from theoretical physics and general relativity research that began appearing in astro-ph through cross-listing.

Over the past decade (2015-2025), 190 new concepts emerged (Appendix~\ref{app:recent_concepts}, Tables~\ref{tab:recent_concepts_part1} and \ref{tab:recent_concepts_part2}). Deep learning dominates recent emergence: Astronomical CNN Applications (1,676 papers), Deep Learning in Astronomy (604 papers), Residual Neural Networks (402 papers), U-Net Variants in Astronomy (373 papers), Transformer Architectures in Astronomy (185 papers), and Physics-Informed Neural Networks (120 papers). Recent concepts also include observational capabilities—including JWST Deep Extragalactic Surveys (75 papers, emerged 2022) and GW170817 Multimessenger Merger (324 papers, emerged 2017), which by our metric are considered "new" when they first crossed the 5-paper threshold, even though JWST's scientific impact continues to grow. Appendix~\ref{app:recent_concepts} provides representative examples from the full list.

\begin{figure*}[t]
  \centering
  \includegraphics[width=0.95\textwidth]{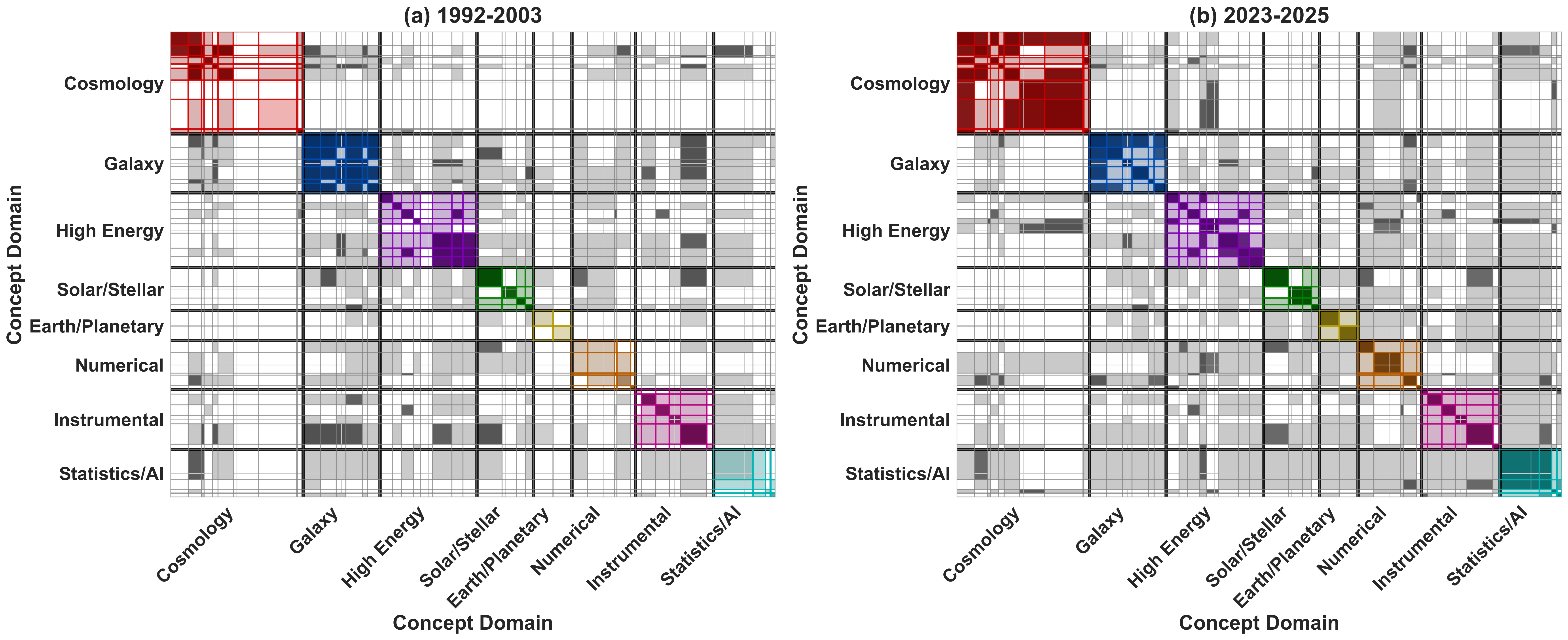}
  \caption{\textbf{Evolution of concept co-occurrence in astrophysics.} Darker colors indicate stronger co-occurrence. (a) Early period (1992--2003): established domain structure. (b) Recent period (2023--2025): computational domains (Statistics/AI, Numerical Simulation) show increased internal coherence and enhanced cross-domain integration with traditional astrophysical domains, reflecting the field's evolution toward data-intensive research.}
  \label{fig:cooccurrence_evolution}
\end{figure*}

\subsection{Concept Co-occurrence}

While temporal analysis reveals when individual concepts emerge, understanding how concepts appear together in papers provides complementary insights into the thematic structure of research. Unlike traditional citation analysis which tracks paper-to-paper relationships, concept co-occurrence reveals how different methodologies, observations, and theories interconnect within the field, identifying which ideas are commonly explored together and how these patterns shift as the field develops.

We quantify co-occurrence using the Ochiai coefficient, which normalizes by concept popularity. Intuitively, if two concepts $i$ and $j$ appear together in $N_{ij}$ papers, and appear individually in $N_i$ and $N_j$ papers respectively, the Ochiai coefficient is:
\begin{equation}
    \text{Ochiai}(i,j) = \frac{N_{ij}}{\sqrt{N_i \cdot N_j}}
\end{equation}
This normalization is important because different subfields have vastly different publication volumes—this ensures we measure genuine conceptual relationships rather than simply reflecting which fields are most active.

Co-occurrence analysis is a rich topic with many dimensions to explore. Here we present a simple comparison between two time periods to illustrate the utility of our concept vocabulary. Figure~\ref{fig:cooccurrence_evolution} compares the earliest window (1992--2003, 40,000 papers) with the most recent window (2023--2025, 40,000 papers). Using fixed-size temporal windows removes field growth bias—later periods do not appear artificially stronger simply due to increased publication volume. 

For visualization, we apply spectral clustering within each of the 8 predefined domains (Table~\ref{tab:categories}) using 2025 data to identify subclusters, producing the fine-grained structure visible in Figure~\ref{fig:cooccurrence_evolution}. To aggregate the 9,999$\times$9,999 concept matrix into this manageable visualization, we compute the 10th percentile of co-occurrence scores within each subcluster block (capturing robust signal while filtering noise), and use the spread between 10th and 30th percentiles to set transparency (indicating consistency of patterns). These percentile choices enhance dynamic range: the 10th percentile provides a stable color metric that is less sensitive to outliers than the median, while the 30th-10th spread reveals whether co-occurrence within a block is consistent (low spread, high transparency) or heterogeneous (high spread, lower transparency). This hierarchical structure is held fixed across all temporal windows, enabling direct comparison.

As shown in the figure, the technical domains—Statistics/AI, Numerical Simulation, and to some extent Instrumentation—exhibit more cross-domain interactions in the recent period compared to the early period. In the recent period, the Statistics/AI domain shows prevalent integration with all astrophysical domains, reflecting the widespread adoption of machine learning and data-driven methods across subdisciplines. The Numerical Simulation domain displays increased internal coherence, consistent with the field's growing reliance on computational methods. These patterns show that computational and statistical approaches have evolved from peripheral tools to core components of the research ecosystem.

Concepts in science domains (Galaxy Physics, High Energy, Solar/Stellar) maintain relatively stable internal structure and interdomain connections across both periods. The Cosmology domain shows notable internal growth along with increased cross-connections to High Energy. This growth is partly attributable to the 2007 cross-listing policy expansion discussed previously, which brought theoretical physics concepts into astro-ph. The Earth/Planetary domain shows increased internal coherence in the recent period, consistent with the expansion of exoplanet research enabled by missions such as \textit{Kepler} and \textit{TESS} in recent years.

This analysis demonstrates how our concept vocabulary enables quantitative study of field evolution in ways that would be difficult or impossible with traditional keyword systems. The patterns revealed—computational integration, methodological shifts, and domain stability—provide empirical evidence for narratives about how astrophysics research has changed over three decades. Appendix~\ref{app:subcluster_cooccurrence} provides representative examples of within-domain and cross-domain concept pairs with strong co-occurrence, demonstrating fine-grained thematic structure. More sophisticated temporal analyses are beyond the scope of this paper, but the released dataset supports such investigations.

\section{Dataset Release}

We release the dataset on GitHub at \url{https://github.com/tingyuansen/astro-ph_knowledge_graph} which covers all astro-ph papers from 1992 through July 2025. The public release prioritizes the concept vocabulary and embeddings to enable reproducibility and support downstream applications. For structured summaries, we adopt a more conservative distribution policy detailed in Appendix~\ref{app:example_summary}. The public release includes: \textbf{Concept vocabulary} as CSV with labels, names, classes, and descriptions; \textbf{concept embeddings} using text-embedding-3-large; \textbf{paper metadata} including year, arXiv ID, and ADS bibcodes; and Python scripts for data loading, verification, and analysis. A complementary \textbf{citation network} extracted from NASA ADS API is also provided, with 1.67M unique identifiers covering both internal references (between astro-ph papers) and external citations (to other disciplines). Table~\ref{tab:stats} summarizes the dataset statistics.

\section{Conclusion}

This work presents a dataset of 408,590 astrophysics papers from arXiv astro-ph (1992-2025) with structured six-section summaries, 9,999 AI-generated concepts with detailed descriptions, and semantic embeddings.

The key contribution is a systematically generated concept vocabulary that addresses limitations of traditional keyword systems. Unlike author-supplied ADS keywords that suffer from extreme sparsity and frequency imbalances, our AI-generated concepts provide consistent coverage across all papers with balanced distributions. Each concept includes a detailed description that preserves scientific context, enabling more effective discovery than single-word keywords. Our embedding space analysis demonstrates that concepts capture dispersed semantic information within papers that abstracts alone cannot represent, making them critical for scientific discovery rather than merely navigation.

\begin{table}[ht!]
  \centering
  \small
  \begin{tabular}{lr}
  \toprule
  \textbf{Metric} & \textbf{Value} \\
  \midrule
  Total papers & 408,590 \\
  Unique concepts & 9,999 \\
  Total concept associations & 3,827,232 \\
  Avg concepts per paper & 9.4 (median: 10) \\
  Avg papers per concept & 383 (median: 223) \\
  \bottomrule
  \end{tabular}
  \caption{Summary statistics of the astro-ph knowledge graph dataset (Table 1). All 408,590 papers have complete structured summaries, concept associations, and semantic embeddings.}
  \label{tab:stats}
\end{table}

Temporal analysis reveals how the concept vocabulary captures field evolution. Recent emergence (2015-2025, 190 concepts, Tables~\ref{tab:recent_concepts_part1} and \ref{tab:recent_concepts_part2}) is dominated by machine learning adoption, while also tracking major observational facilities and theoretical developments. Co-occurrence analysis demonstrates the increasing integration of computational domains (Statistics/AI, Numerical Simulation) with traditional astrophysical research areas, revealing the field's evolution toward data-intensive methodologies. These analyses show the vocabulary's ability to capture both enduring foundations and emerging research frontiers across three decades of astrophysics.

This dataset enables applications including semantic search systems, research trend analysis, knowledge graph construction, and training language models for scientific understanding. The combination of structured summaries, comprehensive concept vocabulary, and semantic embeddings makes this resource suitable for advancing AI-assisted scientific discovery. Recent work has demonstrated the potential of LLM agents in astronomical analysis~\citep{Sun2025,Wang2025b}, and our structured representations provide the foundation for developing autonomous systems in astronomical research.

While this paper focuses on dataset creation and preliminary analysis, extrinsic evaluation through task-based applications is an important next step. We are actively exploring integration with recommender systems and semantic search platforms to enable concept-based paper discovery and citation network analysis. Such applications will provide quantitative evaluation of utility through user studies in production environments.

\clearpage 

\section*{Code and Data Availability}

All code, system prompts, and data processing pipelines are publicly available at \url{https://github.com/tingyuansen/astro-ph_knowledge_graph}. This includes OCR processing scripts, multi-stage summarization prompts, concept extraction and clustering code, embedding generation, co-occurrence calculation, and data verification scripts. While the proprietary APIs are not open-source, all prompts and processing logic are fully documented to enable replication with alternative models.

\section*{Acknowledgments}

YST is supported by the National Science Foundation under Grant AST-2406729.

This research was supported by the NSF National Artificial Intelligence Research Resource Program, the Microsoft Accelerating Foundation Models Academic Research Grant, the OpenAI Research Access Program, and the NVIDIA Academic Grant Program. 

This research used resources of the Oak Ridge Leadership Computing Facility at the Oak Ridge National Laboratory, which is supported by the Office of Science of the U.S. Department of Energy under Contract No. DE-AC05-00OR22725. Additional computational resources were provided by the Australian National Computational Infrastructure.

\bibliography{manuscript}

\clearpage  

\appendix

\section{Example Structured Summary}
\label{app:example_summary}

This appendix provides a representative example of our structured summaries to demonstrate their comprehensive nature and systematic organization. Unlike traditional abstracts that prioritize brevity, our summaries (typically 600-900 words, averaging ~740 words) systematically separate six semantic sections: Background (observational and theoretical context), Motivation (scientific questions and goals), Methodology (technical approach and data), Results (empirical findings), Interpretation (theoretical analysis), and Implication (broader significance). This structure enables targeted information retrieval—a researcher can directly access methodological details or theoretical interpretations without reading the entire paper. While paper summarization is now routine with LLMs, we provide these structured summaries upon request rather than through public release out of caution at this scale (0.4M papers). Researchers interested in accessing the summaries should contact the authors.

\vspace{10pt}

\noindent\rule{\columnwidth}{0.4pt}
\vspace{4pt}
\noindent
\textbf{Title:} Usco1606-1935: An unusually wide low-mass triple system? \\
\textbf{Author:} Adam L. Kraus et al. (arXiv:0704.0455)

\vspace{2pt}
\textbf{Background:} The study of multiple star systems is crucial for understanding star formation processes. Surveys have shown that binary frequencies and properties vary significantly with mass. Solar-mass stars exhibit high binary frequencies ($>$60\%) and can have separations up to $\sim$10$^4$ AU. In contrast, M dwarfs have lower frequencies (30-40\%) and fewer companions beyond $\sim$500 AU, while brown dwarfs show even lower frequencies ($\sim$15\%) with few companions exceeding 20 AU. The observed decline in maximum binary separation with decreasing mass has been described by empirical functions, suggesting that this limit is established early in stellar lifetimes. Surveys of young stellar associations have identified a few unusually wide systems, but not enough to analyze their properties statistically.

\vspace{1pt}
\textbf{Motivation:} To address the scarcity of unusually wide low-mass systems, we utilized archival 2MASS data to search for candidate wide binary systems among known members of three nearby young associations, including Upper Sco. Our findings aim to align with the standard paradigm, revealing a deficit of wide systems among very low-mass stars and brown dwarfs, while also identifying a few candidates, such as USco1606-1935, a wide pair of stars with similar fluxes and colors. This study seeks to evaluate the probability of USco1606-1935 being an unusually wide, low-mass binary, thereby contributing to the understanding of multiple system formation and evolution in young stellar associations.

\vspace{1pt}
\textbf{Methodology:} We identified USco1606-1935 AB as a candidate binary using 2MASS data, leveraging its bright and resolved components to gather additional photometry and astrometry from various surveys, including DENIS, USNO-B, and SSS. The analysis focused on 2MASS $JHK$ magnitudes and USNO-B $I$ magnitudes, ensuring consistency through comparisons with DENIS data... Optical spectroscopy was conducted using the Double Spectrograph at Palomar Observatory, processing the spectrum with standard IRAF tasks and comparing it with spectral standards from Upper Sco and Taurus to confirm the spectral type. High-resolution imaging was achieved with laser guide star adaptive optics on the Keck-II telescope, obtaining nearly diffraction-limited images in both narrow and wide camera modes to measure photometry and astrometry for the components.

\vspace{1pt}
\textbf{Results:} High-resolution images revealed that USco1606-1935 A comprises two sources, Aa and Ab, with the probability of an unbound bright source near A being extremely low, suggesting that Aa and Ab form a bound binary system. Photometric data confirmed that USco1606-1935 B aligns with known members of Upper Sco, supporting its membership, although its position in color-magnitude diagrams raised questions about potential differential reddening or unresolved companions. Astrometric analysis summarized the relative positions of the system components and field stars, with proper motion indicating that B is likely a comoving member... Stellar and binary properties for the Aa-Ab and A-B systems were estimated using isochrones and temperature scales to derive component masses and spectral types.

\vspace{1pt}
\textbf{Interpretation:} Identifying pre-main sequence binaries presents challenges in distinguishing gravitationally bound pairs from coeval, comoving stars. To assess clustering among PMS stars, we calculated the two-point correlation function (TPCF), which quantifies the number of excess pairs at a given separation compared to a random distribution. Utilizing a Monte Carlo approach, the TPCF revealed significant clustering of stars at small separations. The analysis indicated that the expected surface density of unbound companions is $\sim$60 deg$^{-2}$, suggesting a 25\% chance of chance alignments among low-mass members. Consequently, while the detection of Aa and Ab as a close binary is highly probable, the physical association of Aab and B cannot be assumed based solely on probabilistic grounds, highlighting the complexity in confirming wide binary status in such systems.

\vspace{1pt}
\textbf{Implication:} If Aab and B are gravitationally bound, USco1606-1935 would represent one of the rare young multiple systems with wide separations comparable to field systems of similar mass. However, the significant probability of chance alignment necessitates caution in such classifications. This underscores the need for systematic searches for wide binaries in the Upper Sco association and similar young stellar environments to better understand the frequency and properties of wide, low-mass multiple systems. Enhanced observational strategies and comprehensive data analyses are essential to distinguish truly bound systems from coincidental alignments, thereby refining our knowledge of star formation and the dynamical evolution of multiple star systems.
\vspace{4pt}
\noindent\rule{\columnwidth}{0.4pt}

\section{Recent Concept Emergence (2015-2025)}
\label{app:recent_concepts}

A total of 190 concepts emerged during 2015-2025, defined as crossing the 5-paper publication threshold during these years. This represents approximately 2\% of our total vocabulary, reflecting the maturation of the field where new research increasingly builds connections between established concepts rather than introducing entirely new topics. Tables~\ref{tab:recent_concepts_part1} and \ref{tab:recent_concepts_part2} present representative examples from this emergence, focusing on concepts that reflect genuine recent developments in astrophysics methodology, observations, and theory.

The dominance of machine learning and deep learning concepts (46 concepts emerged in 2015 alone) reflects the rapid adoption of AI methods across astrophysics during this period. Traditional methodologies like Monte Carlo simulations and N-body dynamics had already been well-established in the 1990s, but their application within modern neural network architectures represents a distinct conceptual development. The examples shown capture major observational events (GW170817 Multimessenger Merger in 2017) and the scientific impact of new facilities (JWST Deep Extragalactic Surveys in 2022, Gaia-Sausage-Enceladus Merger in 2018).

The declining number of new concepts in very recent years (2 in 2025, 4 in 2024, 7 in 2023) reflects several factors discussed in the main text. First, many fields have matured, with research increasingly focused on connections between established concepts rather than entirely new topics. Second, our 5-paper threshold means concepts can appear earlier than their peak importance—papers from 2024-2025 have had less time to accumulate the required citations. Third, our clustering methodology may exhibit systematic bias: genuinely novel concepts appearing in recent years may be merged into established clusters from earlier periods if sufficiently similar in embedding space. However, the continued emergence of new concepts demonstrates that even mature fields continue generating new research directions.

\begin{table*}[t]
\centering
\scriptsize
\begin{tabular}{lr|lr}
\toprule
\textbf{Concept} & \textbf{Papers} & \textbf{Concept} & \textbf{Papers} \\
\midrule
\multicolumn{4}{c}{\textbf{2015}} \\
\midrule
Astronomical CNN Applications & 1676 & Extremely Randomized Trees & 50 \\
Deep Learning in Astronomy & 604 & Odd Radio Circles & 46 \\
Astronomical Data Augmentation & 403 & Planetary Similarity Metrics & 41 \\
Autoencoder Architectures & 281 & Global 21-cm Signal & 38 \\
Astronomical Transfer Learning & 273 & Planetary Weather Simulation Systems & 37 \\
Exoplanet Atmospheric Retrieval Systems & 162 & MeerKAT Data Pipelines & 32 \\
Cosmic Reionization Simulations & 144 & CMB Interaction Effects & 25 \\
Precision-Recall Evaluation & 127 & Millimeter-Wave Technology Integration & 22 \\
Skill Score Metrics & 108 & Protostellar Evolutionary Metrics & 18 \\
Rapid Bayesian Sky Localization & 84 & Gravitational Wave Data Systems & 17 \\
Astronomical Anomaly Detection Pipelines & 54 & Plasma Momentum Dynamics & 15 \\
Nuclear Matter Meta-Modeling & 54 & CORDIC-based Signal Processing & 13 \\
Detection Metric Balance & 53 & Nonlinear Supersymmetry and Gravity Theories & 9 \\
\midrule
\multicolumn{4}{c}{\textbf{2016}} \\
\midrule
Gravitational Wave Mergers & 202 & Planetary Robotic Mobility Systems & 43 \\
Exoplanet Radiative Transfer Codes & 198 & Non-Minimal Coupling Models & 31 \\
Recurrent Neural Networks & 169 & Trust Region Optimization Methods & 29 \\
t-SNE and Topological Data Analysis & 157 & OPTICS Clustering Techniques & 28 \\
No-U-Turn Sampling & 117 & Low-Noise Transistor Technologies & 26 \\
Astronomical Classification Techniques & 106 & Snow Uncertainty Mitigation in Ice Detection & 26 \\
Synthetic Minority Oversampling & 88 & Infrared Stellar Outbursts & 22 \\
Data-Driven Spectral Inference & 86 & Asteroid Exploration Missions & 21 \\
Joule-Thomson Thermodynamics & 77 & Solar ALMA Integration & 14 \\
Sub-Threshold Signal Analysis & 57 &  &  \\
Continuous Wave Detection Algorithms & 54 &  &  \\
Mars Atmospheric and Thermal Studies & 44 &  &  \\
\midrule
\multicolumn{4}{c}{\textbf{2017}} \\
\midrule
Residual Neural Networks & 402 & Interstellar Object Dynamics & 100 \\
GW170817 Multimessenger Merger & 324 & Probabilistic Neural Networks & 100 \\
S8 Clustering Discrepancy & 240 & Batch Normalization in Neural Networks & 78 \\
Adversarial Neural Architectures & 196 & Thermal Protection Systems & 34 \\
DHOST Theories & 167 & Titan Aeolian Dynamics Exploration & 29 \\
Deep Learning Frameworks & 159 & SOXS Optical and Control Architecture & 23 \\
Graph Neural Networks in Astronomy & 121 &  &  \\
Electron Lepton Number Dynamics & 102 &  &  \\
Kilonova Emission Modeling & 101 &  &  \\
\midrule
\multicolumn{4}{c}{\textbf{2018}} \\
\midrule
U-Net Variants in Astronomy & 373 & Rapid Blue Transients & 41 \\
Gaia-Sausage-Enceladus Merger & 215 & Particle Spray Simulation & 38 \\
LSTM Neural Architectures & 160 & Dirac-Fermion Stars & 37 \\
PHANGS Astronomical Surveys & 118 & FLASK Cosmological Simulation and Web Framework & 31 \\
Inception-Based Neural Networks & 96 & Protoplanetary Disk Substructure Research & 29 \\
EDGES 21-cm Anomaly & 56 & SPHINX Cosmological Simulations & 28 \\
Astronomical Data Sonification & 52 &  &  \\
Interpretable Machine Learning in Astronomy & 49 &  &  \\
CubeSat Scientific Missions & 43 &  &  \\
Remote Sensing Indices and Nighttime Imaging & 42 &  &  \\
\bottomrule
\end{tabular}
\caption{Recent concept emergence (2015-2018): Part 1 showing representative examples. Concepts sorted by total papers within each year.}
\label{tab:recent_concepts_part1}
\end{table*}
\begin{table*}[t]
\centering
\scriptsize
\begin{tabular}{lr|lr}
\toprule
\textbf{Concept} & \textbf{Papers} & \textbf{Concept} & \textbf{Papers} \\
\midrule
\multicolumn{4}{c}{\textbf{2019}} \\
\midrule
Probabilistic Transformation Flows & 288 & VGG-based Neural Networks & 63 \\
Neural Inference Methods & 202 & SH0ES Hubble Constant Measurement & 45 \\
Variational Autoencoders & 176 & Dataset Tension and Suspiciousness Metrics & 37 \\
Quantum Entanglement Islands & 139 & Neutrino Event Reconstruction Methods & 30 \\
Physics-Informed Neural Networks & 120 & Primordial Black Hole Dynamics & 26 \\
Explainable AI Visualization Techniques & 79 & Lyman-Alpha Tomography & 26 \\
Deep Learning for Astronomy & 79 & Atmospheric Refraction and Polarimetry Models & 11 \\
Astronomy-Focused AI Language Models & 69 & Helium Suppression Phenomena & 8 \\
Commensal Radio Astronomy Surveys & 67 &  &  \\
\midrule
\multicolumn{4}{c}{\textbf{2020}} \\
\midrule
Advanced Attention Mechanisms & 154 & Gaia Black Hole Binaries & 22 \\
Barrow Entropy in Cosmology & 77 & ALMA Protoplanetary Chemistry Studies & 15 \\
Yebes 40m QUIJOTE Survey & 76 & Seismic Noise Mitigation for Gravitational Observatories & 11 \\
Satellite Brightness Mitigation & 61 &  &  \\
Bern Planetary Formation Model & 25 &  &  \\
\midrule
\multicolumn{4}{c}{\textbf{2021}} \\
\midrule
Transformer Architectures in Astronomy & 185 & T-ReX Cosmic Analysis & 19 \\
Astronomical Image Datasets & 36 & YSO Characterization Techniques & 9 \\
Photon Propagation Simulations & 21 &  &  \\
\midrule
\multicolumn{4}{c}{\textbf{2022}} \\
\midrule
JWST Deep Extragalactic Surveys & 75 & Cosmology Data Efficiency Techniques & 13 \\
Lyman-Alpha Forest Correlations & 19 & Stingray Astrophysical Analysis & 11 \\
Lorentz Violation in High-Energy Phenomena & 13 &  &  \\
\midrule
\multicolumn{4}{c}{\textbf{2023}} \\
\midrule
Astrochemical Molecular Analysis & 8 & Pulsar Signal Analysis Methods & 6 \\
\midrule
\multicolumn{4}{c}{\textbf{2024}} \\
\midrule
Adaptive Neural Architectures & 17 & Galactic Foreground Contamination & 10 \\
Distributed Sampling Efficiency & 12 &  &  \\
\midrule
\multicolumn{4}{c}{\textbf{2025}} \\
\midrule
Rotating Outflow Dynamics & 6 &  &  \\
\bottomrule
\end{tabular}
\caption{Recent concept emergence (2019-2025): Part 2 showing representative examples.}
\label{tab:recent_concepts_part2}
\end{table*}

\section{Subcluster Co-occurrence Patterns}
\label{app:subcluster_cooccurrence}

The co-occurrence analysis in Section 6 reveals fine-grained substructure within each primary domain. Table~\ref{tab:subcluster_examples} presents representative concept pairs exhibiting strong co-occurrence within domains and across domain boundaries, illustrating the thematic patterns visible in Figure~\ref{fig:cooccurrence_evolution}. The Ochiai coefficients quantify co-occurrence strength normalized by concept frequency.

These patterns demonstrate the rich thematic structure within the concept vocabulary. Within-domain pairs reveal specialized research areas: cosmological theories (axion-like particles, Bianchi models), AGN dynamics (reverberation mapping, episodic jets), stellar physics (sunspot dynamics, variable stars), and computational methods (molecular spectroscopy, hydrodynamic simulations). Cross-domain pairs reveal methodological connections: cosmological dynamics linking with numerical stability analysis, radiative transfer simulations connecting Galaxy Physics with Numerical methods, neutrino and gamma-ray detection bridging High Energy physics with specialized instrumentation, helioseismology connecting Solar physics with time-series analysis, and gravitational wave template matching linking Numerical simulations with statistical inference methods.

\begin{table*}[t]
\centering
\scriptsize
\begin{tabular}{lllc}
\toprule
\textbf{Domain Pair} & \textbf{Concept 1} & \textbf{Concept 2} & \textbf{Ochiai} \\
\midrule
\multicolumn{4}{c}{\textit{Within-Domain Co-occurrence}} \\
\midrule
Cosmology \& Nongalactic & Axion-Like Particle Phenomenon & Photon-ALP Oscillations & 0.594 \\
& Complexity-Volume Conjecture & Holographic Complexity & 0.538 \\
& Anisotropic Cosmology & Bianchi Cosmological Models & 0.420 \\
& Einstein-Cartan Theories & Spacetime Torsion Dynamics & 0.406 \\
\midrule
Galaxy Physics & AGN Reverberation Mapping & Broad-Line Region Dynamics & 0.377 \\
& Double-Double Radio Galaxies & Episodic AGN Jet Activity & 0.368 \\
& Galactic Pattern Speeds & Tremaine-Weinberg Methods & 0.360 \\
& Quasar Broad Absorption Dynamics & Quasar Outflow Dynamics & 0.347 \\
\midrule
High Energy Astrophysics & Superhump Dynamics & SU UMa-Type Dwarf Nova Superoutbursts & 0.560 \\
& Double Degenerate SN Progenitors & Single Degenerate SN Progenitors & 0.455 \\
& GZK Cosmic Ray Limit & Ultra-High Energy Cosmic Rays & 0.447 \\
& Black Hole Entropy Dynamics & Quantum Entanglement Islands & 0.434 \\
\midrule
Solar \& Stellar Physics & Sunspot Flow Dynamics & Sunspot Penumbra Dynamics & 0.516 \\
& Cepheid Variable Stars & Variable Star Distance Scaling & 0.421 \\
& Blazhko Effect Dynamics & RR Lyrae Stars & 0.386 \\
& Standard Solar Model & Solar Neutrino Dynamics & 0.378 \\
\midrule
Earth \& Planetary Science & Light Pollution Dynamics & Night Sky Brightness Quantification & 0.562 \\
& Extraterrestrial Signal Assessment & Technosignature Detection & 0.450 \\
& Graphene Curvature Dynamics & Graphene Quantum Analogues & 0.428 \\
& Geomagnetic Activity Metrics & Geomagnetic Storm Dynamics & 0.404 \\
\midrule
Numerical Simulation & Molecular Dipole Moments & Molecular Spectroscopy Computation & 0.317 \\
& Molecular Spectroscopy Computation & Partition Functions in Astrophysics & 0.292 \\
& Astrophysical Hydrodynamic Simulations & FARGO Numerical Simulation Suite & 0.280 \\
& Potential Energy Surfaces & Quantum Coupled Interactions & 0.279 \\
\midrule
Instrumental Design & Acoustic Neutrino Detection & Underwater Acoustic Positioning Systems & 0.368 \\
& Satellite Brightness Mitigation & Satellite Astronomical Interference & 0.353 \\
& Atmospheric Seeing Instrumentation & Atmospheric Turbulence Dynamics & 0.348 \\
& Axion Haloscope Detection & Resonant Cavity Systems & 0.333 \\
\midrule
AI/Statistics & Neural Inference Methods & Simulation-Based Inference & 0.491 \\
& Transformer Architectures in Astronomy & Advanced Attention Mechanisms & 0.432 \\
& Nonextensive Statistical Mechanics & Nonextensive Tsallis Thermodynamics & 0.343 \\
& Astronomical CNN Applications & Astronomical Data Augmentation & 0.296 \\
\midrule
\multicolumn{4}{c}{\textit{Cross-Domain Co-occurrence}} \\
\midrule
Cosmology $\leftrightarrow$ Numerical Simulation & Cosmological Dynamical Systems & Fixed and Critical Points Stability & 0.417 \\
& Poisson Sprinkling in Causal Sets & Causal Set Quantum Gravity & 0.375 \\
& Fuzzy Dark Matter Mechanics & Schrödinger-Poisson Dynamics & 0.346 \\
& Bose-Einstein Condensate Phenomena & Gross-Pitaevskii-Poisson Dynamics & 0.309 \\
\midrule
Galaxy $\leftrightarrow$ Numerical Simulation & Sersic Light Distribution & Galaxy Modeling Software & 0.208 \\
& Lyman Alpha Line Profiles & Lyman Alpha Radiative Transfer & 0.204 \\
& Ionization State Dynamics & Photoionization Models & 0.201 \\
& Gas-Grain Surface Chemistry & Astrochemical Modeling Systems & 0.196 \\
\midrule
High Energy $\leftrightarrow$ Instrumental & High-Energy Cosmic Neutrinos & IceCube Neutrino Observatory & 0.403 \\
& Black Hole Shadow Phenomenon & Global Interferometric BH Imaging & 0.340 \\
& Very High Energy Gamma Rays & Imaging Atmospheric Cherenkov Telescopes & 0.259 \\
& Cosmic Ray Air Showers & Cosmic Ray Radio Detection & 0.255 \\
\midrule
Solar/Stellar $\leftrightarrow$ AI/Statistics & Skill Score Metrics & Solar Cycle and Flare Prediction & 0.302 \\
& Helioseismic Signal Correlations & Helioseismic Travel-Time Kernels & 0.296 \\
& Stellar Flare Frequency Dynamics & Automated Flare Detection & 0.279 \\
& Mass-to-Flux Ratio Dynamics & Davis-Chandrasekhar-Fermi Method & 0.233 \\
\midrule
Earth/Planetary $\leftrightarrow$ Instrumental & Meteor Stream Dynamics & Global Meteor Observation Networks & 0.358 \\
& VLF/ULF Electromagnetic Phenomena & VLF Electromagnetic Observation Systems & 0.357 \\
& Mesospheric Sodium Layer Dynamics & Guide Stars in Adaptive Optics & 0.293 \\
& Meteoroid Trajectory Analysis & Global Meteor Observation Networks & 0.274 \\
\midrule
Numerical $\leftrightarrow$ AI/Statistics & Cellular Automaton Systems & Self-Organized Criticality & 0.239 \\
& Gravitational Wave Template Banks & Gravitational Wave Matched Filtering & 0.213 \\
& Kernel-Based Seismic Inversion & Regularized Inversion Methods & 0.180 \\
& Poincar\'e Analysis & Lyapunov Measures in Chaos & 0.172 \\
\bottomrule
\end{tabular}
\caption{Representative concept co-occurrence patterns within and across primary domains. Within-domain pairs show specialized research themes with multiple representative examples per domain from actual co-occurrence analysis. Cross-domain pairs reveal methodological connections between fields, including the integration of computational and statistical methods with traditional astrophysics domains. All pairs extracted from empirical co-occurrence across 408,590 papers (1992-2025) using Ochiai normalization.}
\label{tab:subcluster_examples}
\end{table*}

\end{document}